\documentclass{aastex}

\newcommand{\vdag}{(v)^\dagger}
\newcommand{\myemail}{demers@astro.umontreal.ca}

\slugcomment{Submitted to the Astronomical Journal}

\shorttitle{2MASS observations of C stars}
\shortauthors{Demers et al.}

\begin{document}


\title{2MASS observations of spectroscopically identified extragalactic C stars}


\author{Serge Demers and Mathieu Dallaire}
\affil{D\'epartement de Physique Universit\'e de Montr\'eal, Montreal, Qc
H3C 3J7, Canada}
\email{demers@astro.umontreal.ca; tnc@videotron.ca}

\and
\author{Paolo Battinelli}
\affil{Osservatorio Astronomico di Roma, viale del Parco Mellini 84, 
I-00136 Roma, Italy}
\email{battinel@oarhp1.rm.astro.it}


\begin{abstract}
We matched spectroscopically identified C stars (from low resolution 
objective prism surveys) 
in the Magellanic Clouds
with 2MASS sources.  We confirm that C stars
show a large spread in absolute magnitudes, even in the  K$_s$ band. 
We show that the I and K$_s$ magnitude distributions 
of a population of C stars (in the LMC) have a similar narrow dispersion if
the C stars are selected in a well defined color range. 
Using magnitude and color criteria, we employ the 2MASS data to identify  
26 C stars in the the Fornax dwarf spheroidal galaxy.

The mean K$_s$ magnitude and the mean bolometric magnitude 
of C stars are found to be slightly brighter in the LMC and SMC when
compared to those of the Fornax dwarf spheroidal galaxy. 
The difference could be explained by ages or/and abundance differences.
\end{abstract}



\section{Introduction}
Thanks to the 2MASS survey (Skrutskie et al. 1997) the astronomical community
has now J, H, and K$_s$ magnitudes for almost 
all the brighter stars of the Magellanic
Clouds and a number of nearby galaxies, since their (S/N = 10) magnitude
limits are respectively 16.3, 15.3 and 14.7. These observations are particularly
suitable to investigate the red giant branch (RGB) and asymptotic giant branch (AGB)
populations of nearby galaxies. Because the Magellanic Clouds are nearby, stars of
fainter absolute magnitudes have been acquired by 2MASS. Indeed, Nikolaev \& Weinberg
(2000) have used the 2MASS data to investigate the spatial 
distribution of several types of stars of the Large Magellanic Cloud (LMC). The
various types of stars were defined by their location on the 
K$_s$ vs (J -- K$_s$) color-magnitude diagram.  

What is of immediate interest to us is their J region which contains 
carbon-rich AGB stars and is defined by a sample of C-rich long period variables
from Hughes \& Wood (1990).  Nikolaev \& Weinberg (2000) state that stars in this
region of the CMD are potentially good standard candles because their K$_s$ magnitude
spread is quite small, providing that stars are selected in a narrow color range.
This important aspect has been followed up by Weinberg \& Nikolaev (2001) to probe the
three-dimensional structure of the LMC. They, however, restrict even more the 
color interval of the stars in the J region by selecting only stars in 
the range 1.6 $<$ J -- K$_s$ $<$ 1.7. 

This approach is quite impressive, but it can be applied only when thousands
of stars are available since the narrow color range reduces appreciably
the size of the sample.
The question we ask then, is how useful will
be the K$_s$ magnitudes of C stars if one were to select stars in the J
region (1.4 $<$ J--K$_s$ $<$ 2.0) rather than the narrower color range.
The goal of this approach is to investigate C star populations in other
nearby galaxies where dozens of C stars are seen rather than thousands.

\section{The C star samples}

We search the 2MASS database to obtain magnitudes of two sets of 
spectroscopically identified C stars. By {\it spectroscopically identified} we
mean low resolution slitless objective prism spectroscopy. Such surveys in 
crowded fields produce spurious identification for faint objects. They are: 

1-- Set {\bf KDMK}: Kontizas et al. (2001) recently published a catalogue of 7760
C stars in the LMC. The stars were identified from an objective prism survey of
the LMC with the UK Schmidt Telescope. The authors provide RI photographic 
magnitudes
obtained from the UK Schmidt plates. Their list includes hundreds of stars,
observed in RI, by Costa \& Frogel (1996). The comparison of the magnitudes
and colors reveals that it is quite difficult to achieve any accuracy
with  UK Schmidt photographic photometry. We will not use these magnitudes
and colors.

2-- Set {\bf MH}: 1185 C stars in $\sim 220$ deg$^2$ covering the SMC and the 
inter-cloud region are identified by Morgan \& Hatzidimitriou (1995) from 
UK Schmidt Telescope objective-prism plates. Neither magnitudes nor colors 
are available for these stars.

Two other sets of Magellanic Cloud C stars were considered but not retained, 
they are: the list of C stars in the periphery of the LMC and SMC 
spectroscopically identified by 
 Kunkel, Irwin \& Demers (1997) and Kunkel, Demers \& Irwin (2000). Nearly all
of these stars are included in the LMC KDMK and the MH SMC sets. 
Costa \& Frogel (1996) have
published I and (R--I) for hundreds of C stars in the LMC. 
Their stars were selected from  lists of objective prism spectral 
identifications by Blanco et al. (1980) and
Blanco \& McCarthy (1983). Unfortunately, these stars have rather 
poorly established equatorial coordinates.  
Cross identifications with 2MASS sources lead to matches
as far off as 30 arcsec from the best 2MASS candidates. For this reason
we prefer not to use directly the original coordinates. Most of these
stars are expected to be in the KDMK set but they were not all cross
identified by Kontizas et al. (2001).

\section{Color-magnitude diagrams}

7078 stars  of the KDMK set were matched, within 2 arcsec of 2MASS sources. 
These stars are displayed on the color magnitude diagram in Figure 1.
It is obvious that spectroscopically identified C stars possess a wide
range of K$_s$ magnitudes and (J--K$_s$) colors. This figure
 shows that C stars are
found not only on the AGB tip but also at lower magnitudes on the AGB 
which nearly coincides, in magnitudes and colors, with the giant branch. 
The parallelogram outlines the J region, defined by  Nikolaev \& Weinberg
(2000) where C stars have a small range of K$_s$ magnitudes. We adopt
its color boundaries as: 1.4 $<$ J--K $<$ 2.0, the upper and lower
magnitude boundaries depend on the distance of the C stars.
 
\placefigure{fig1}
The variable color excess within the LMC and the effect of the depth of the LMC
on the apparent magnitudes increase the K$_s$ dispersion of C stars. 
The East - West inclination
of the disc of the LMC introduces a $\pm 0.2$ apparent magnitude dispersion
(Nikolaev \& Weinberg 2000).
The bulk of the C stars are near the center/bar, thus only the relatively few
in the periphery would contribute to the widening of the distribution. 
The extinction, A$_K$, is at most a few hundredth of a magnitude (Nikolaev \&
Weinberg 2000). Cioni et al. (2000) adopt A$_K$ = 0.04 for both Magellanic Clouds.
 The K$_s$ magnitude distribution of the 7078 LMC C stars
is shown in Figure 2. The shade histogram corresponds to stars with J--K colors
in the 1.4 to 2.0 range. 
Obviously, the K$_s$ magnitudes are better defined by selecting C stars in a narrow
color interval; the FWHM of the shaded distribution is 0.33 mag. 
No correction has been
made for the color excess. Since E(J--K) = 0.65E(B-V), the reddening corrections
are expected to be $\approx 0.07$ mag.
\placefigure{fig2}
\placefigure{fig3}
The MH set yields 1093 matches. The CMD of this set is presented in Figure 3. As in
the LMC case, we observe that spectroscopically identified C stars show a large
range of K$_s$ magnitudes. 
Figure 4 compares the magnitude
distributions of the LMC and SMC C stars located in the J region.  The
average total reddening toward the two Magellanic Clouds is nearly
similar (Westerlund 1997). The mean
magnitude of the distributions being respectively 10.68 and 11.26 
for a difference of
0.58 mag. 
\placefigure{fig4}
\section{Color-color diagrams}
The photometric properties of C stars in the color-color diagram have
been investigated by Cohen et al. (1981). They found slight differences, 
explained
by the different metallicities between the LMC, SMC and Galactic C star
populations. Figure 5 displays the color-color diagram of the LMC C stars.
The two solid lines outline the J--K limits of the J region. Ninety nine percent
of the stars in the J region are within the box limited by the two dashed
lines. 
\placefigure{fig5}
The SMC C stars are displayed on the color-color diagram in Figure 6, 
the same box than Fig. 5 is
shown. One notes that SMC C stars are indeed 
slightly displaced in colors relative to the LMC ones. 
\placefigure{fig6} 
\section{DISCUSSION}
\subsection{Comparison of $\langle I\rangle$ and $\langle K_s\rangle$}
Albert, Demers \& Kunkel (2000) and Letarte et al. (2002) have shown that, when a large
C star population is observed in a galaxy, the C star
M$_I$ magnitude distribution has a narrow FWHM. 
We are now in position to compare the
dispersion of I and K$_s$ magnitudes of C stars. To do so we have, however,
to photometrically select a sub-sample of C stars with well defined colors
and not simply accept all
spectroscopically defined C stars. We already know that, for all C stars, there
is a substantial magnitude spread. For the purpose of our comparison we
need C stars with I and K$_s$ magnitudes. 
Costa \& Frogel (1996) have obtained RI magnitudes of some 800  spectroscopically
identified C stars in the LMC.  
Following our adopted color criterion, we select only C stars with
(R--I)$_o$ $>$ 0.90, see for example,
Albert et al. (2000); Battinelli \& Demers (2000) or the corresponding V--I
criterion of Brewer, Richer \& Crabtree (1996). 
Some 300 C stars satisfying this criterion have been observed by
Costa \& Frogel (1996) and are cross identified 
in the KDMK data set. We thus have also
the K$_s$ magnitude of these stars. It is rather interesting to note
that the R--I color criterion is rather similar to  the J region
criterion in the K$_s$ vs J--K$_s$ plane. Indeed, on Figure 7 we plot
the 300 LMC C stars used for our comparison. Eighty seven percent of the
stars are in the J region.
\placefigure{fig7}
The comparison of the I and K$_s$ magnitude distributions, shown in
Figure 8, reveals that C stars, selected as described above, have a 
{\it narrower} I$_o$ magnitude distribution than K$_s$. One must however
take note that Costa \& Frogel (1996) have corrected their magnitudes
for reddening while the K$_s$ are not corrected. They adopt a mean A$_I$ for 
a given area, based on the extinction of clusters. Since the reddening was
not determined for each star, one must consider it approximate. We redetermined the
I$_o$ magnitude distribution by adding to the magnitudes a random reddening variation ranging
up to E(R--I) = $\pm$ 0.05, corresponding to a maximum A$_I$ = $\pm$ 0.12, values
quite reasonable for the LMC. The only
effect of this extinction variation 
is to increase slightly the width of the magnitude distritution, keeping it 
narrower than the K$_s$ magnitude distribution. The A$_K$ variation
across the face of the LMC is expected to be small and probably does
not explain the larger width of the distribution. 
>From the two mean magnitudes,  $\langle I_o\rangle$ = 13.8 and $\langle K_s\rangle$ = 10.6 we
see that C stars are 3.2 magnitudes brighter in K$_s$ than in I. 
It does not follow, however, that K$_s$ observations of C stars are more
time efficient than I observations. The sky is so much brighter in the K
band than the I band than to obtain similar S/N, a much longer 
exposure time is required in K$_s$ than in the I band. 
\placefigure{fig8}
\subsection{The 2MASS survey toward Fornax}
The Fornax dwarf spheroidal galaxy, the most massive of the dwarf
spheroidal associated with the Milky Way has been known for over twenty
years to contain an intermediate age population and carbon stars 
(Demers \& Kunkel 1979; Aaronson \& Mould 1980). During the last decades, 
spectroscopic surveys have permitted the identification of dozens of
C stars in Fornax.
Azzopardi (1999) mentions that there are 104 known C stars but barely
half of these have published coordinates, magnitudes or colors. 
\placefigure{fig9}
We have found 4365 2MASS sources within one degree from the center of
Fornax and with a signal in the three bands. 
The color-magnitude diagram of these sources, presented in Figure 9,
reveals that only a few stars are have the right color to be C stars. 
Taking, (m--M)$_o$ = 20.76 for the true modulus of Fornax 
(Demers, Kunkel
\& Grondin 1990; Saviane, Held \& Bertelli 2000), 
one would expect C stars to have K$_s \approx 13$. 
Indeed, inspection of the J--K$_s$ color errors as a function of the
apparent magnitudes shows that for K$_s > 14.0$ photometric errors
become larger than $\pm$0.1 mag. 
This implies that stars seen on the CMD near the magnitude limit 
have poorly determined magnitudes and colors.
We therefore select only stars, in the J region with photometric 
errors less than 0.1 mag as possibly Fornax C star candidates. 
The stars in the J region, which satisfy the above magnitude criterion
fall in the expected rectangle in the color-color diagram, shown in 
Figure 10.  
\placefigure{fig10}
We have thus identify, using JHK$_s$ photometry, 
 26 C stars in Fornax. They represent a sub-sample of the C star population
of Fornax because we select stars in a well defined 
color range. These C stars are given in Table 1. Listed are their
Equatorial J2000 coordinates, and their 2MASS magnitudes and colors. 
Cross identifications with  previously
known C stars or known red stars are given 
in Table 2. We include here V, B--V (photographic magnitudes) from
Demers, Irwin \& Kunkel (1994) when available.
Most of the 26 C stars  are already
spectroscopically identified C stars. Seven of them are newly
confirmed C stars, including two not found in our database of red 
stars in Fornax. The newly identified C stars are in the periphery of
Fornax, outside of the regions previously surveyed by low dispersion
spectroscopy by Frogel et al. (1982) and Westerlund et al. (1987).
\placetable{tab-1}
\placetable{tab-2}
The mean K$_s$ magnitude of the 26 identified C stars of Fornax is
$\langle K_s\rangle$= 13.08, corresponding to $\langle M_{K_s}\rangle = -7.68$  for the  adopted
distance given above. The reddening toward Fornax is quite small,
E(B--V) $\approx 0.03$ thus negligible in the K band.
\subsection{The constancy of $\langle M_{K_s}\rangle$ from galaxy to galaxy.}
Having obtained the mean K$_s$ for the C star population of three galaxies 
we investigate the effect of the metallicity on the mean absolute and
bolometric magnitudes.
We adopt the distances of the LMC and SMC
recently determined, from the tip of the giant branches, by Cioni et al. (2000):
(m--M)$_o$ = 18.55 $\pm 0.04$, for the LMC and (m--M)$_o$ = 18.99 $\pm 0.03$ 
for 
the SMC along with A$_{K_s}$ = 0.04 for both galaxies. Our mean magnitudes
yield:  $\langle M_{K_s}\rangle$ = --7.91
for the LMC and  $\langle M_{K_s}\rangle$ --7.77 for the SMC. The difference between 
these two magnitudes
is larger than the uncertainties on the distances and the reddening. 
\placetable{tab-3}
Table 3 summarizes our findings. The abundance of C stars in the
LMC and SMC are expected to be somewhat less than the canonical metallicities
which correspond to younger populations rather than  the intermediate age 
ones. Intermediate age clusters were found to have [Fe/H] = --0.6, in the
LMC (Rich, Shara, Zurek 2001) and  $-1.6 < [Fe/H] < -1.1$ for the SMC
(Piatti et al. 2001). Table 3 shows that C stars are slightly brighter, in the K$_s$
band, in metal rich systems. The magnitude trend, seen in Table 3,
 is contrary to what Demers \& Battinelli (2002) have observed for the
$\langle M_I\rangle$ of C stars in several Local Group galaxies. In the I band, 
C stars are brighter in the metal poorest systems. 

This effect is, however, not unexpected. Indeed, the AGB tip of
Bertelli's et al. (1994) isochrones is 0.3 mag.  brighter in K for a 
metallicity [Fe/H] = --0.5 than it is for [Fe/H] = --1.5. The AGB tip is
however 0.6 mag. fainter in I for the high metallicity than it is for 
the low metallicity. The absolute magnitude variations are thus confirmed
by the isochrones. These isochrones, however,
reach only J--K = 1.2, a color bluer than the adopted color range of C stars. 

We also transform the K$_s$ magnitudes into bolometric magnitudes using
Frogel, Persson \& Cohen (1980) relation. Nikolaev \& Weinberg (2000)
have shown that the K$_s$ magnitude is essentially equivalent to the K
magnitude. We adopt, for the LMC and SMC a mean
extinction of A$_K$ = 0.04 and reddening of E(J--K) = 0.13. There is 
a slightly larger dispersion among the  mean bolometric magnitudes 
compared to the  mean M$_{K_s}$
magnitudes of the three galaxies listed in Table 3.

Recent isochrones, specifically calculated for carbon stars, by Mouhcine
(2002) reveal that the bolometric magnitudes and M$_{K_s}$  of C stars 
are function
of their age and of their abundance. C stars in the 0.3 to 1 Gyr range
are $\sim 1$ mag. brighter than their older cousins. The brighter
mean M$_{K_s}$ of C stars in the LMC and SMC could then be due to 
the presence of numerous younger C stars, such stars are absent in
Fornax. 

\section{CONCLUSION}
The R--I, CN--TiO approach to identify C stars in galaxies is
equivalent to the  JHK$_s$ photometric technique. The magnitude dispersions
of C stars in K$_s$ and I are about the same. Even if C stars are $\sim 3$
magnitudes brighter in K$_s$ than in I, the sky brightness in K$_s$ is such
that the exposure times in the near infrared are in fact longer than
in R or I. The narrow CN and TiO filters require however much longer exposures
than I or R to reach similar S/N. For this reason the JHK$_s$ photometry
is approximately equivalent in telescope time to the four band photometry.
The  facts that CN and TiO filters are unavailable  in most observatories
and that the new larger telescopes favor near infrared cameras
make the JHK$_s$ approach more promising to survey galaxies in the
neighborhood of the Local Group.


\acknowledgments

This publication makes use of data products from the Two Micron All Sky Survey, 
which is a joint project of the University of Massachusetts and the 
Infrared Processing and Analysis Center, funded by the National Aeronautics and 
Space Administration and the National Science Foundation.
This project is supported financially, in part, by the Natural Sciences and
Engineering Research Council of Canada (S. D.).

\clearpage



\figcaption[demers.fig1.ps]{Color-magnitude diagram of LMC spectroscopically
identified C stars. The parallelogram traces the J region where C stars
with homogeneous properties are found.
\label{fig1}}

\figcaption[demers.fig2.ps]{Magnitude distribution of LMC C stars. The shaded 
histogram represents the C stars in the J region.
\label{fig2}}

\figcaption[demers.fig3.ps]{Color-magnitude diagram of SMC spectroscopically
identified C stars. The J region is lowered by half a magnitude
relative to Figure 1.
\label{fig3}}

\figcaption[demers.fig4.ps]{Comparison of the K$_s$ LMC and SMC magnitude distributions
of C stars in the J region.\label{fig4}}

\figcaption[demers.fig5.ps]{The color-color diagram of the LMC C stars. The solid
lines trace the 1.4 $<$ J--K $<$ 2.0 limits. 99\% of the C stars within
these limits are inside of the dashed lines.\label{fig5}}

\figcaption[demers.fig6.ps]{The color-color diagram of the SMC C stars. The box
is at the same position than on the previous figure. \label{fig6}}

\figcaption[demers.fig7.ps]{Spectroscopically identified C stars with $(R-I)_o > 0.90$.
Nearly all of them fall into the J region.
\label{fig7}}
\clearpage

\figcaption[demers.fig8.ps]{Comparison of the magnitude distribution of a sample
of LMC C stars selected from a R--I color criterion. \label{fig8}}

\figcaption[demers.fig9.ps]{Color-magnitude diagram of the 2MASS sources in
a circle of 1 degree radius centered on the Fornax dwarf spheroidal
galaxy. The color limits of the J region are outlined. \label{fig9}}

\figcaption[demers.fig10.ps]{Stars in the J region outlined in Figure 9
with color errors less than 0.1 mag. fall in the expected region
of C stars. These stars have K$_s$ magnitudes between 12 and 14.
\label{fig10}}


\tablenotetext{a}{Sample footnote for table~\ref{tbl-1} that was generated
with the deluxetable environment}
\tablenotetext{b}{Another sample footnote for table~\ref{tbl-1}}

\tablecomments{Occasionally, authors wish to append a short
paragraph of explanatory notes that pertain to the entire table, but
which are different than the caption.  Such notes should be placed in
a {\tt tablecomments} command like this.}

\clearpage

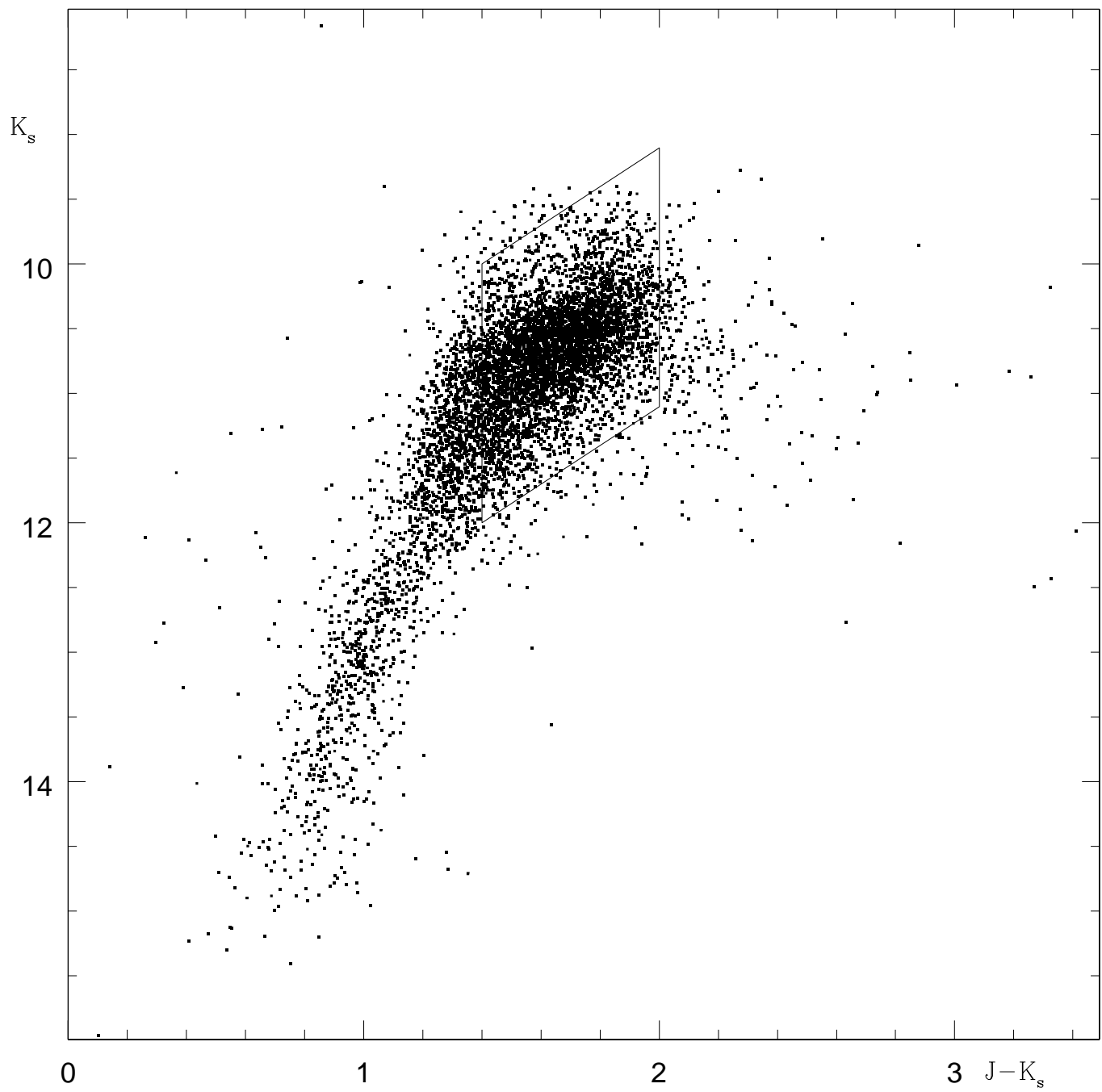
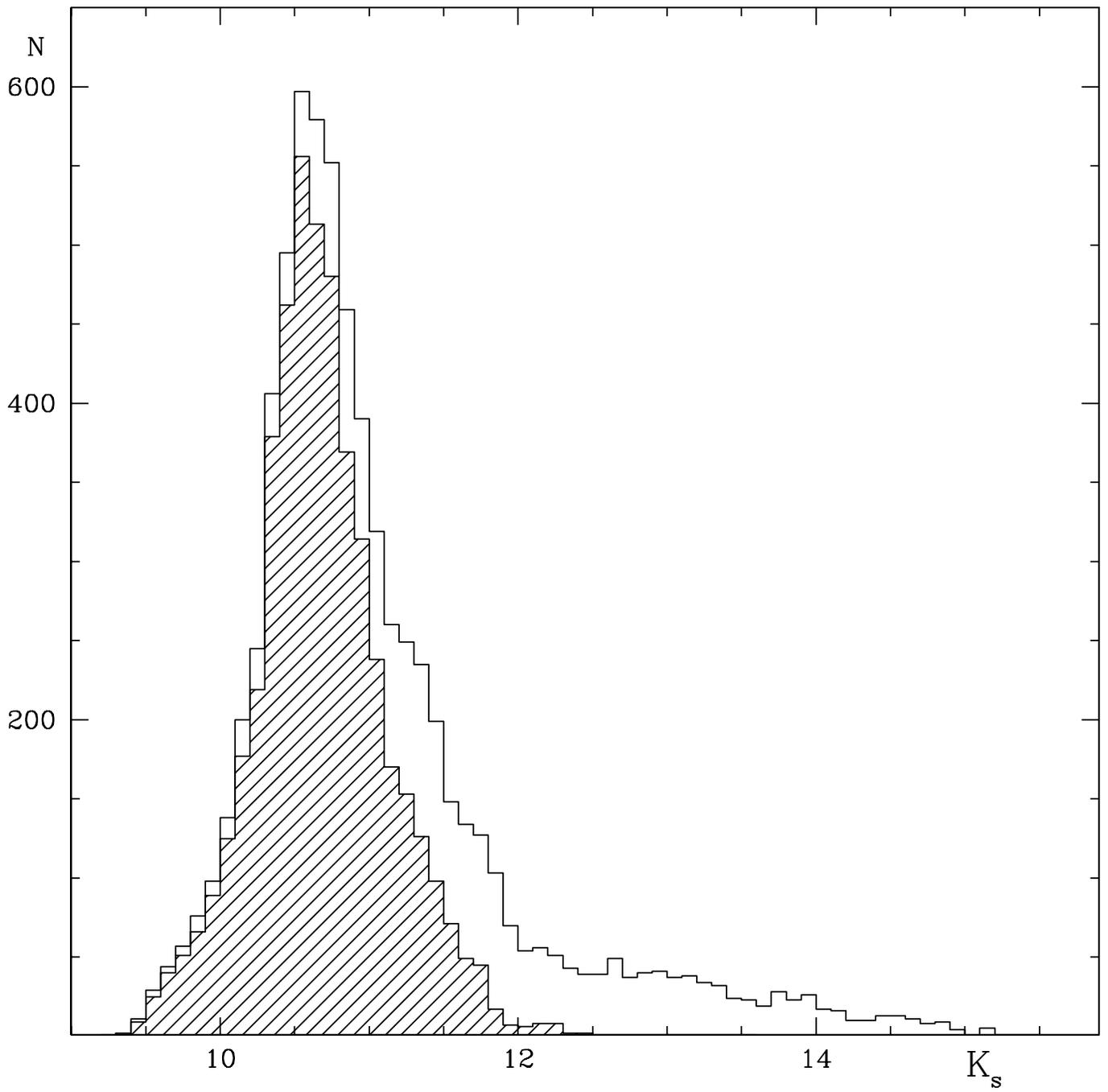
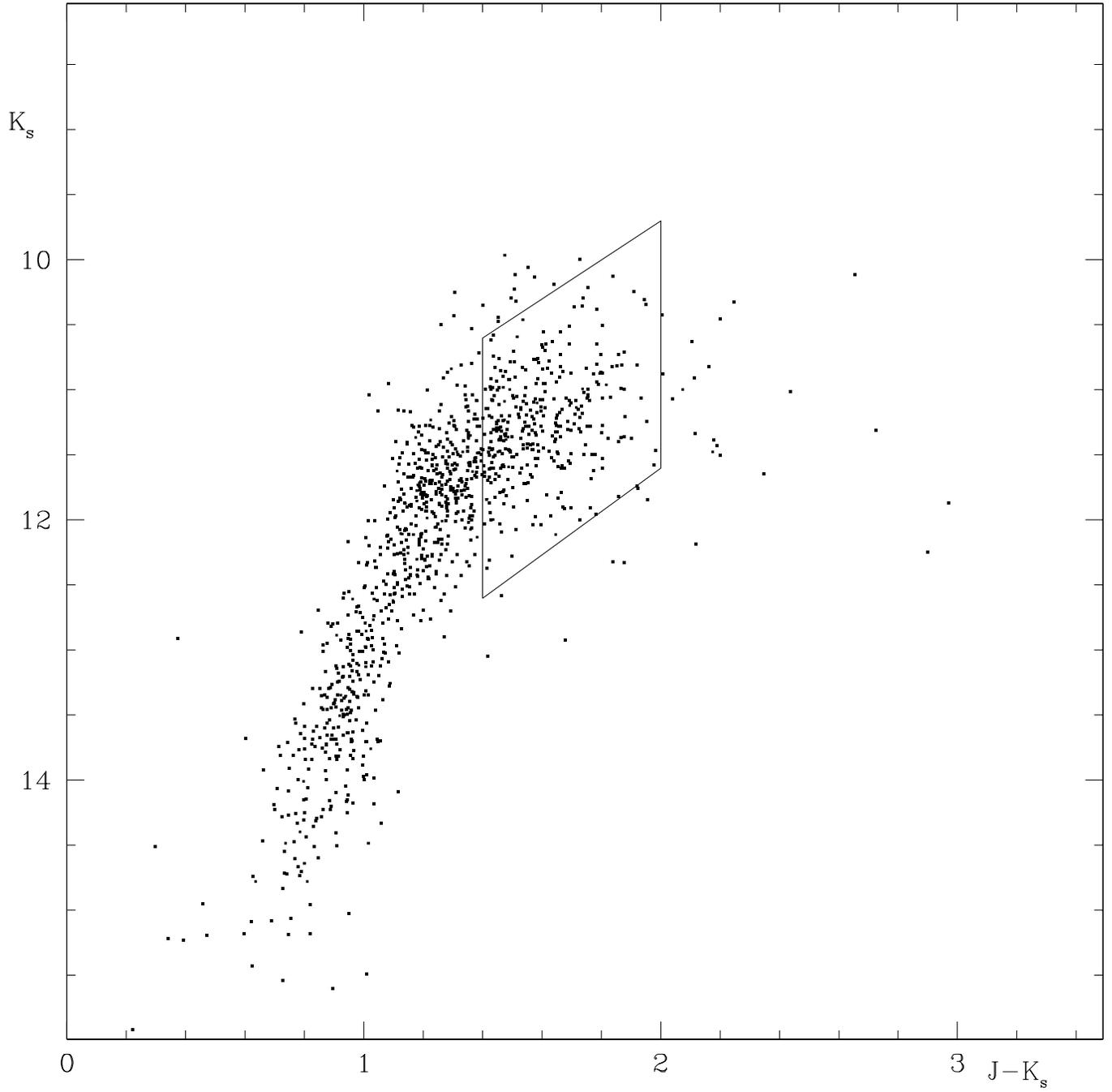
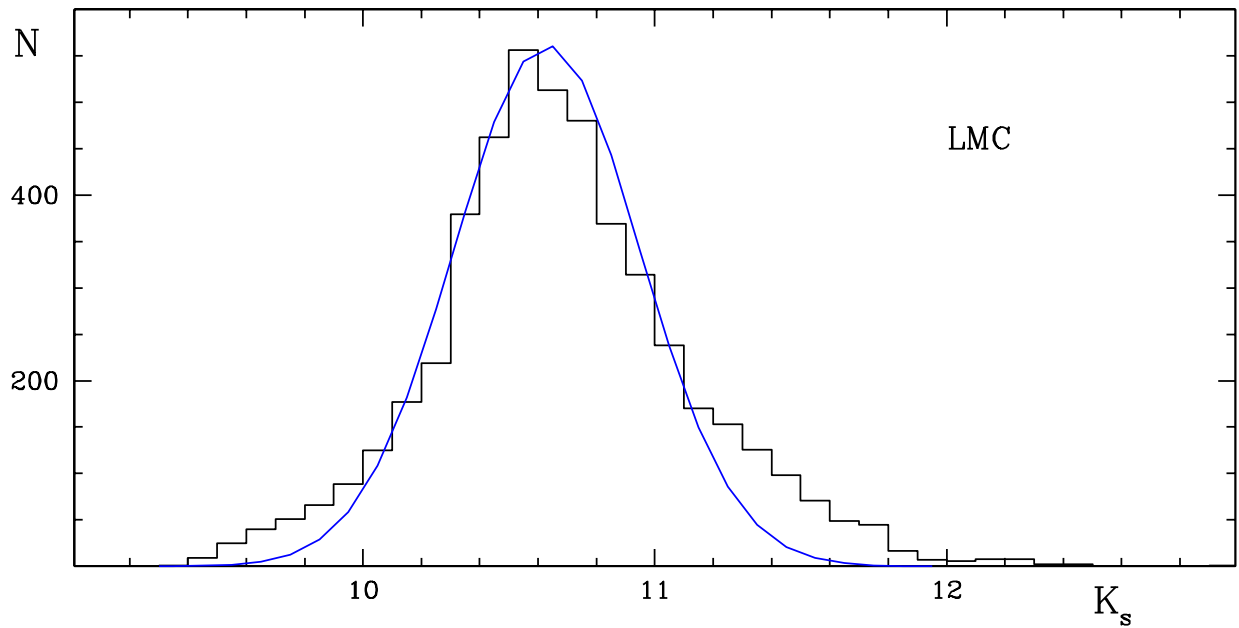
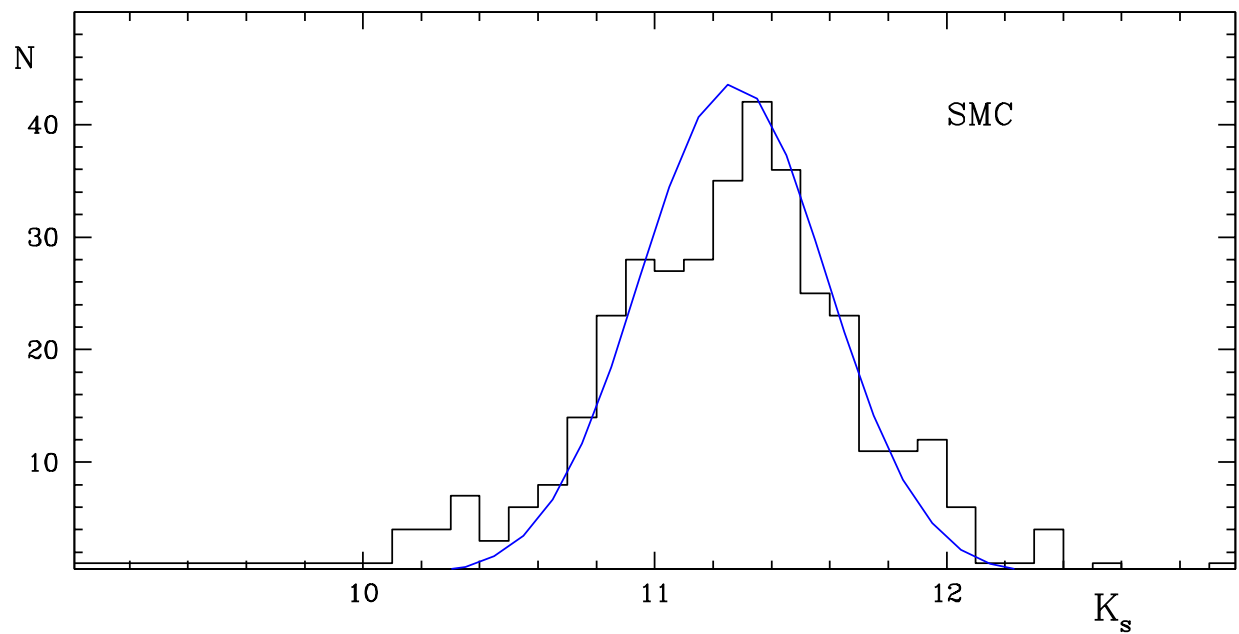
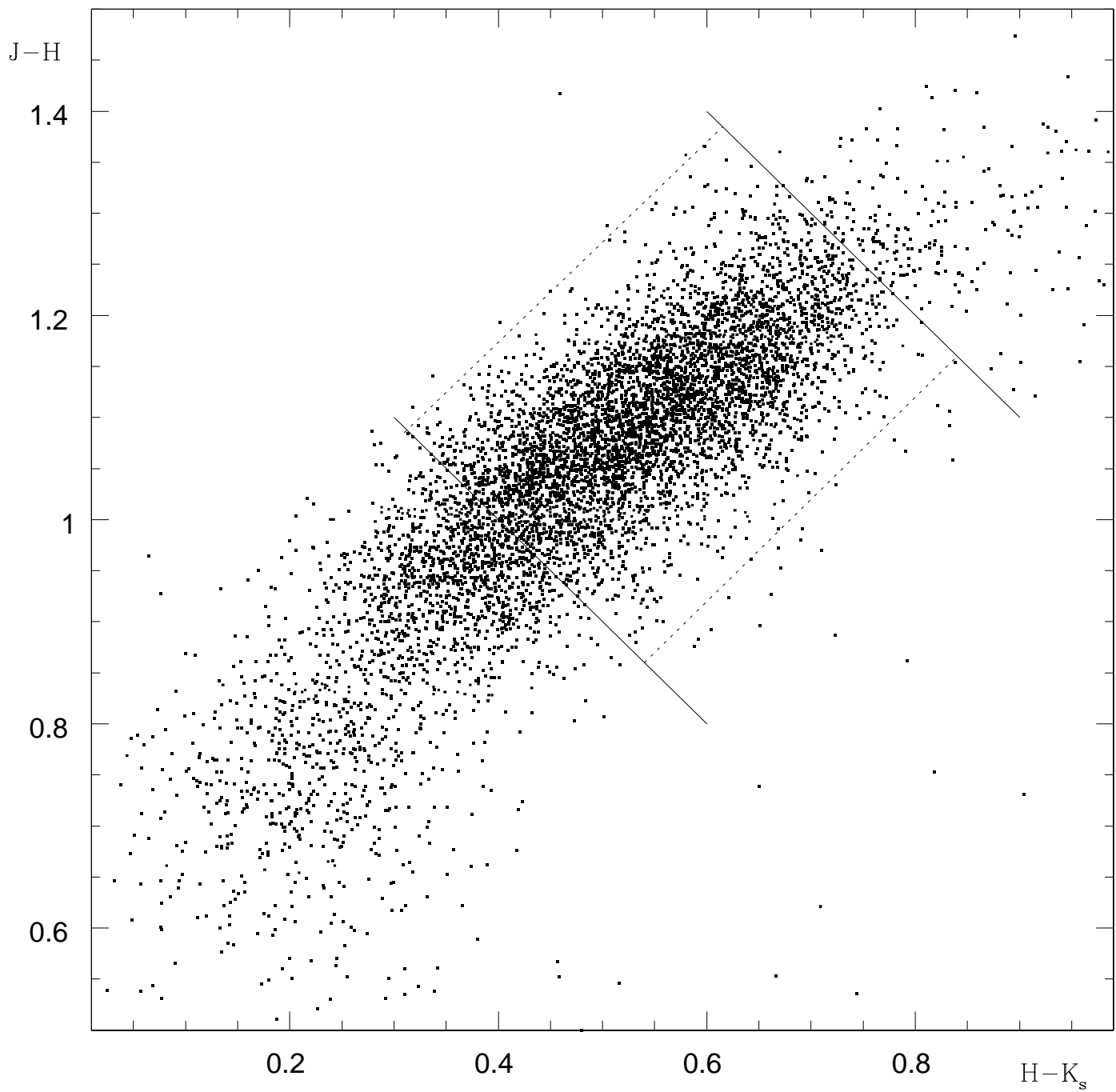
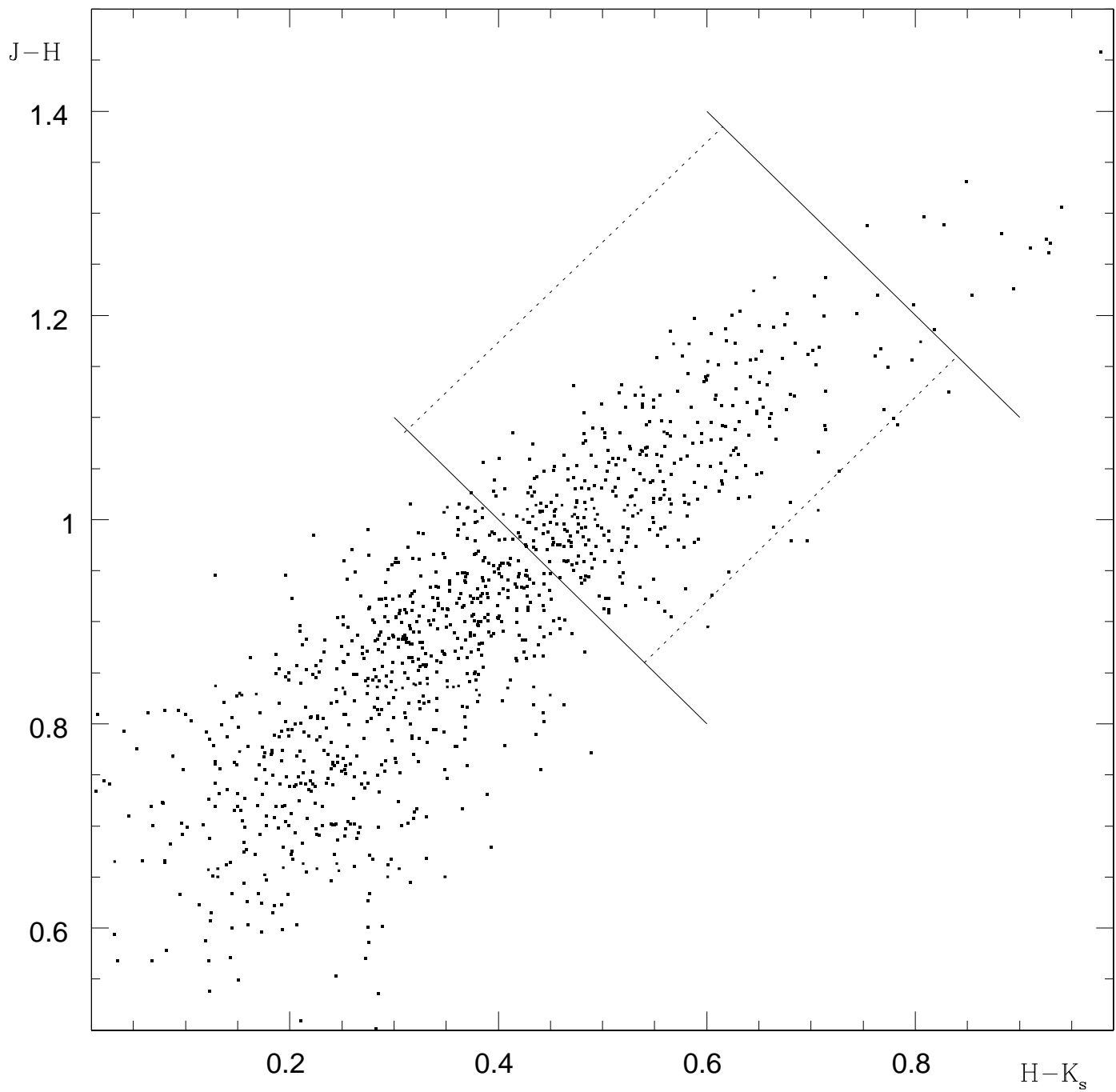
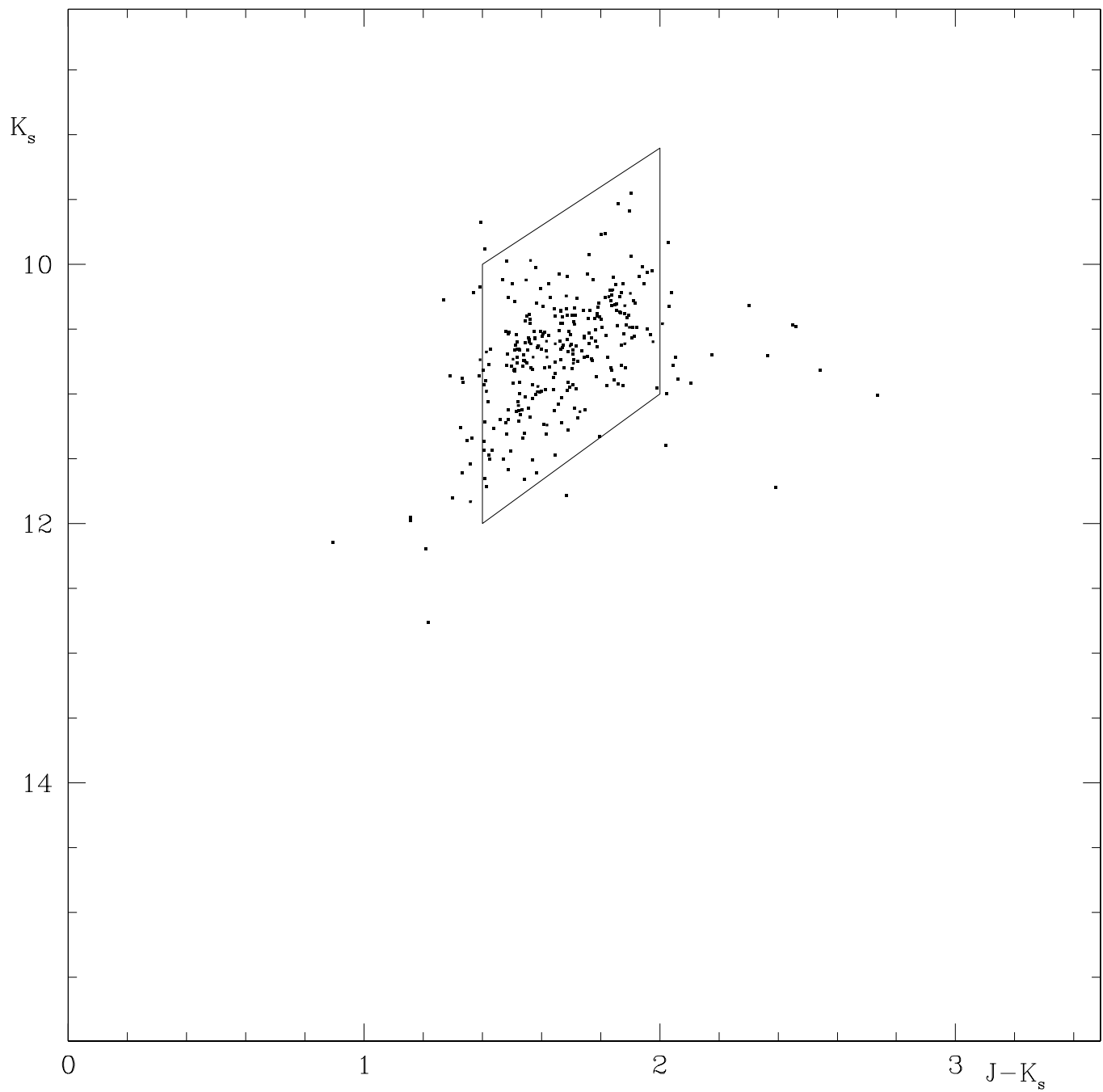
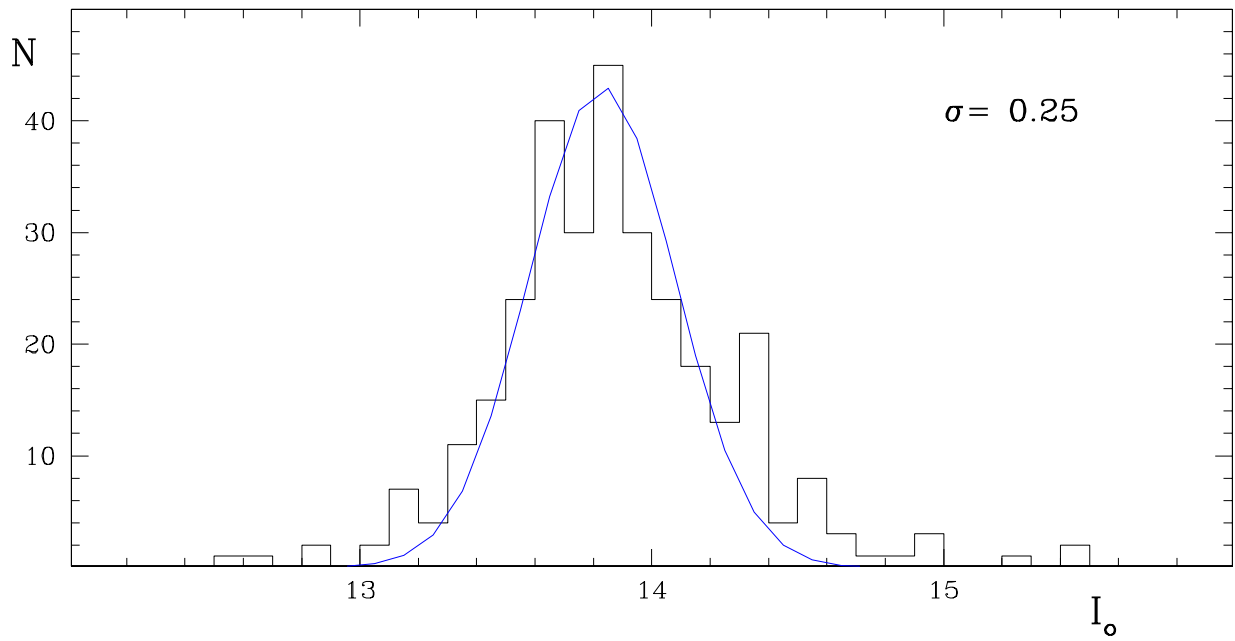
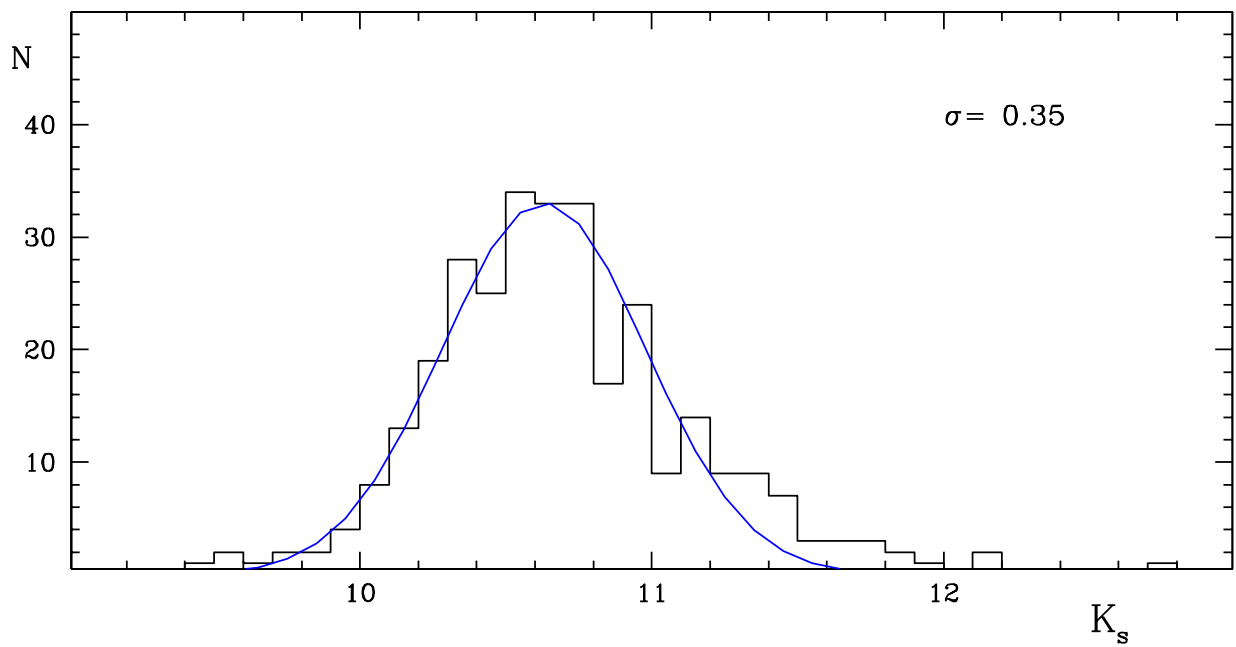
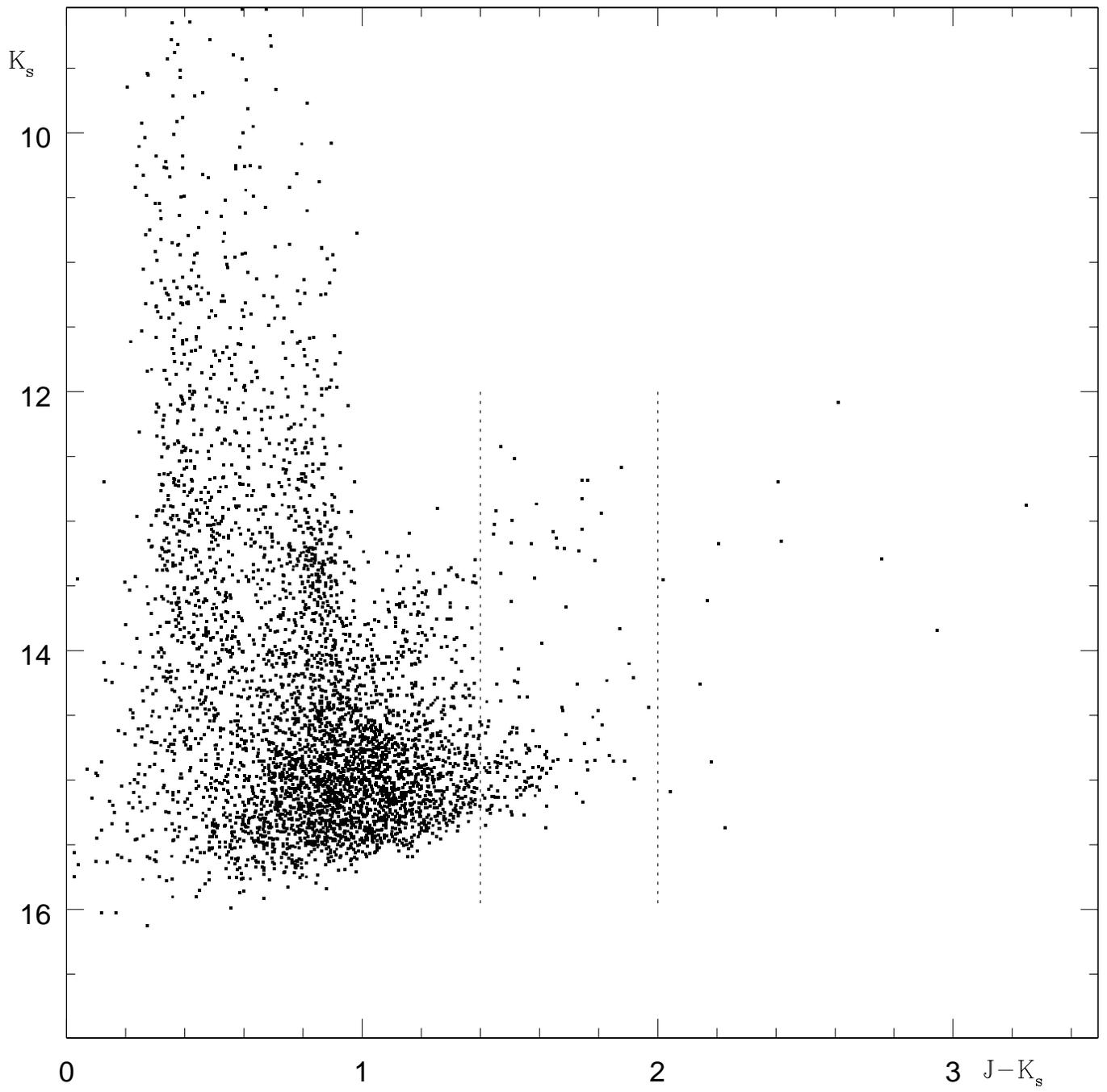
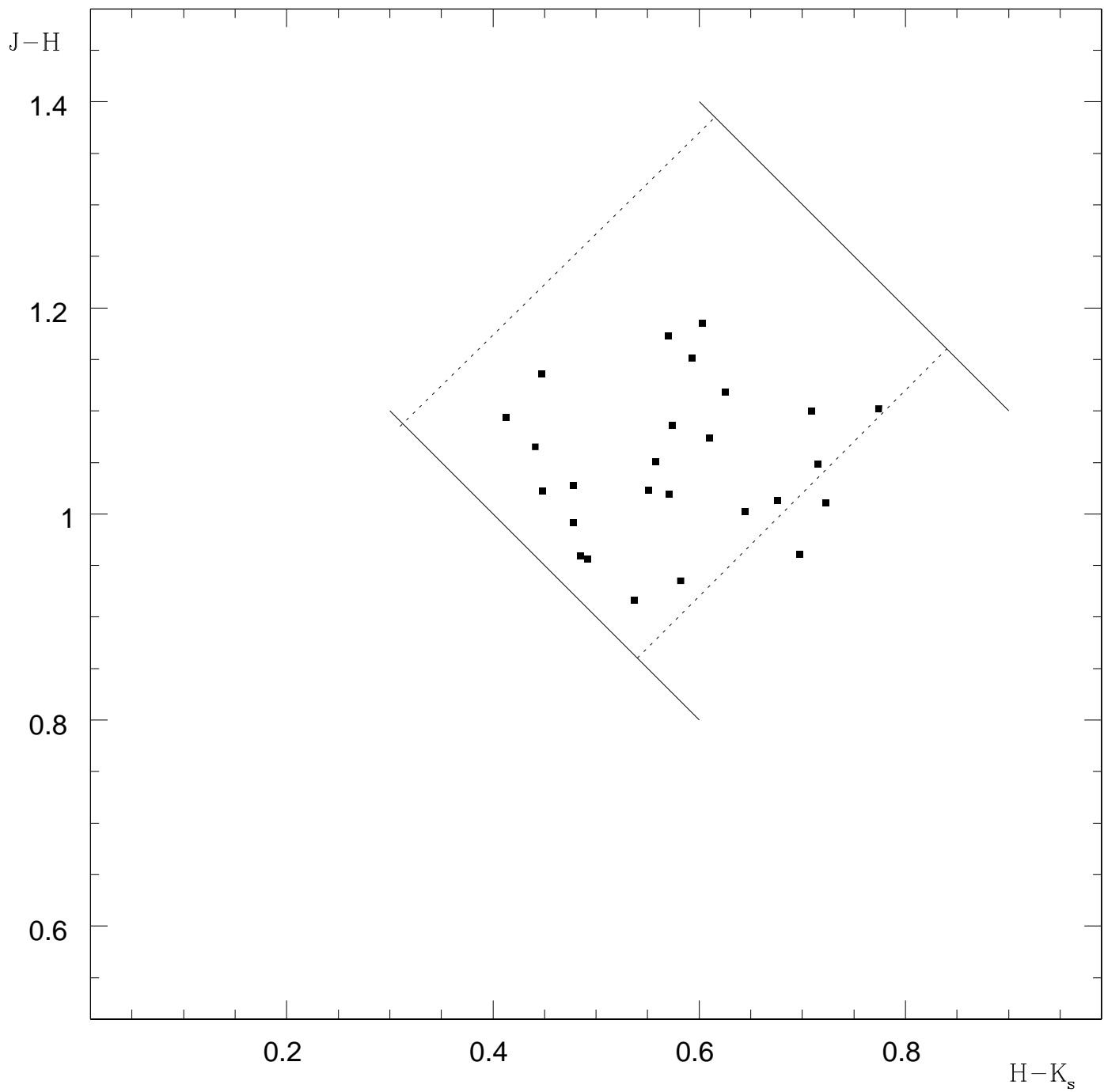
\end{document}